\RequirePackage{rotating}
\documentclass[useAMS,usenatbib]{mn2e}

\def\aap{AA}

\def\apjl{ApJL}

\def\mnras{MNRAS}
\def\apj{ApJ}

\def\aj{AJ}
\def\pasp{PASP}

\def\nat{Nat}

\usepackage{bm}
\usepackage{cancel}
\usepackage{graphicx}
\usepackage{float}
\usepackage{amssymb}
\usepackage{amsfonts}
\usepackage{amsmath} 
\usepackage{color}
\usepackage{natbib}
\usepackage{hyperref}
\usepackage{longtable}
\usepackage{rotating}

\def\ucla{Department of Physics and Astronomy, PAB, 430 Portola Plaza, Box 951547, Los Angeles, CA 90095-1547, USA}
\def\eso{European Southern Observatory, Karl-Schwarzschild-Strasse 2, 85748 Garching bei M{\"u}nchen, DE}
\def\mia{Millennium Institute of Astrophysics, Chile}
\def\unab{Departmento de Ciencias F\'isicas, Universidad Andres Bello, Fernandez Concha 700, Las Condes, Santiago, Chile}
\def\ucd{Department of Physics, University of California, Davis, 1 Shields Avenue, Davis CA 95616, USA}
\def\uv{Instituto de F\'isica y Astronom\'ia, Universidad de Valpara\'iso, Avda. Gran Breta\~na 1111, Playa Ancha, Valpara\'iso 2360102, Chile}

\def\astcam{Institute of Astronomy, University of Cambridge, Madingley Road, Cambridge CB3 0HA, UK}

\def\aaemail{\tt aagnello@eso.org}
\def\pwemail{\tt pwilliams@astro.ucla.edu}

\title[Three new lensed quasars in SDSS]{Discovery of three strongly lensed quasars in the Sloan Digital Sky Survey} 
\author[Williams et al.]{
  Peter R. Williams$^{1,}$\thanks{\pwemail,\aaemail},
  Adriano Agnello$^{2},$
  Tommaso Treu$^{1},$
  Louis E. Abramson$^{1},$\and
  Timo Anguita$^{3,4},$
  Yordanka Apostolovski$^{3,4},$
  Geoff C.-F. Chen$^{5},$\and
  Christopher D. Fassnacht$^{5},$
  J.-W. Hsueh$^{5},$
  Veronica Motta$^{6},$
  Lindsay Oldham$^{7},$\and
  Karina Rojas$^{6},$
  Christian E. Rusu$^{5},$
  Anowar J. Shajib$^{1},$
  Xin Wang$^{1}$
  \medskip\\
  $^1$\ucla\\
  $^2$\eso\\
  $^3$\unab\\
  $^4$\mia\\
  $^5$\ucd\\
  $^6$\uv\\
  $^7$\astcam\\
}

\begin{document}

\voffset-.6in

\date{Accepted . Received }

\pagerange{\pageref{firstpage}--\pageref{lastpage}} 

\maketitle

\label{firstpage}

\begin{abstract}
We present the discovery of 3 quasar lenses in the Sloan Digital Sky Survey (SDSS), selected using two novel photometry-based selection techniques. The J0941+0518 system, with two point sources  separated by 5.46$^{\prime\prime}$ on either side of a galaxy, has source and lens redshifts $z_s = 1.54$ and $z_l = 0.343$. The AO-assisted images of J2211+1929 show two point sources separated by 1.04$^{\prime\prime},$ corresponding to the same quasar at  $z_s = 1.07,$ besides the lens galaxy and Einstein ring. Images of J2257+2349 show two point sources separated by 1.67$^{\prime\prime}$ on either side of an E/S0 galaxy. The extracted spectra show two images of the same quasar at redshift $z_s = 2.10$. In total, the two selection techniques identified 309 lens candidates, including 47 known lenses, and 6 previously ruled out candidates. 55 of the remaining candidates were observed using NIRC2 and ESI at Keck Observatory, EFOSC2 at the ESO-NTT (La Silla), and SAM and the Goodman spectrograph at SOAR. Of the candidates observed, 3 were confirmed as lenses, 36 were ruled out, and 16 remain inconclusive. Taking into account that we recovered known lenses, this gives us a success rate of at least 50/309 (16\%). This initial campaign demonstrates the power of purely photometric selection techniques in finding lensed quasars. Developing and refining these techniques is essential for efficient identification of these rare lenses in ongoing and future photometric surveys.
\end{abstract}

\section{Introduction}
\label{sect:intro}
Strong gravitational lensing serves as a unique probe into the distant universe \citep[e.g.][and references therein]{T+E15}. With their highly magnified images, one can use lenses as cosmic telescopes to study, e.g., the properties of quasar host galaxies at high redshifts \citep{Pen++06qsob,Din++17}. Anomalies in the positions and fluxes of the images can be used to probe dark matter substructure in the lensing object \citep{M+S98,D+K02,Veg++12,Nie++14}. Microlensing due to compact objects in the lens galaxy \citep[see][]{Wamb06} can be used to study the inner structure of lensed AGN, enabling measurements of the accretion disk size \citep{Koch04,Mot++17} and thermal slopes \citep{Ang++08,Eig++08a} and the geometry of the broad line region \citep{Bra++14,Bra++16}. With additional monitoring, one can measure the time delay between arrival of the different images and use this as a cosmological distance indicator \citep[e.g.][]{Ref64,Sch++97,
Suy++14,TCM13,T+M16,BonvinEtal2016}.

Unfortunately, the field is currently limited by the small number of known lenses. Since strong gravitational lensing requires such a close alignment of a distant source with a foreground lensing object, lensed quasars are very rare objects. \citet{O+M10} estimate that, given an $i$-band limiting magnitude of 21, there are only $\sim$0.2 lenses per square degree, of which $\approx$20\% are information-rich quads. Thus, the large footprints of wide-field surveys such as the Sloan Digital Sky Survey \citep[SDSS,][]{Yor++00} and the Dark Energy Survey \citep[DES,][]{DES14} are essential for successful searches.

\begin{table*} 
\centering
\begin{tabular}{lllccc}											
\hline											
Telescope	&	Instrument	&	Type	&	filter/wavelength coverage	&	pixel scale/dispersion	&	slit size ($^{\prime\prime}$)	\\
\hline											
Keck 1	&	OSIRIS	&	Imaging	&	Kbb	&	0.02 arcsec/pixel	&	--	\\
Keck 2	&	NIRC2 narrow	&	Imaging	&	K$^\prime$	&	0.01 arcsec/pixel	&	--	\\
Keck 2	&	ESI	&	Spectroscopy	&	3900 \AA ~to 10900 \AA	&	0.16 to 0.30 \AA/pixel	&	0.75	\\
NTT	&	EFOSC2	&	Spectroscopy	&	3685 \AA ~to 9315 \AA	&	5.54 \AA/pixel	&	1.2	\\
SOAR	&	SAM	&	Imaging	&	SAMI-z	&	0.045 arcsec/pixel	&	--	\\
SOAR	&	Goodman	&	Spectroscopy	&	4912 \AA ~to 9020 \AA	&	1.00 \AA/pixel	&	1.0	\\
\hline											
\end{tabular}											
\caption{Summary of telescopes and instruments used for follow-up of candidates.}
\label{tab:instruments}
\end{table*}

Previous systematic searches for strongly lensed quasars have predominantly explored samples of objects that have spectroscopic data. In the radio, the Cosmic Lens All Sky Survey \citep[CLASS, ][]{Mye++03,Bro++03} in combination with the Jodrell-Bank VLA Astrometric Survey \citep[JVAS, ][]{Kin++99} explored flat-spectrum radio sources, resulting in the discovery of 22 lenses. In the optical, \citet{Pindor:2003p6264} compared fits of single- and double-component point-spread functionss (PSFs) to spectroscopically confirmed quasars in SDSS to identify closely separated pairs of quasars as lens candidates. The SDSS Quasar Lens Search \citep[SQLS, ][]{Og++06,Ina++12} explored the sample of low-redshift ($0.6 < z < 2.2$) spectroscopically confirmed quasars and used a combination of a morphological selection aimed at finding small-separation candidates and a colour-based selection to find lenses that are deblended in SDSS imaging. More recently, \citet{Anu16a} applied a similar method to the Baryon Oscillation Spectroscopic Survey \citep[BOSS, ][]{Daw++13}, expanding the SDSS spectroscopic searches out to higher redshifts.

With the aim of expanding searches to include the footprints of new and upcoming wide-field surveys, many photometry-based techniques have recently been developed. \citet{Fe++16} use Gaussian Mixture Models to search for lenses in the DES Y1A1 \citep{DES14} footprint using DES photometry combined with the Wide-field Infrared Survey Explorer \citep[WISE,][]{WISE10} and VISTA Hemisphere Survey \citep[VHS,][]{VHS13} infrared bands. \citet{Agn++15b} used an artificial neural network classifier applied to blue, extended objects, and \citet{Lin++17} search for red galaxies with multiple blue neighbours in DES.

In this paper, we present the discovery of three new gravitationally lensed quasars selected by two independent photometry-based selection techniques applied to the SDSS DR12 footprint: J0941+0518 at (ra, dec) = (09:41:22.5, +05:18:23.9), J2211+1929 at (22:11:30.3, +19:29:13.2), and J2257+2349 at (22:57:25.4, +23:49:30.4). The quasar images are separated by 5.46$^{\prime\prime}$, 1.04$^{\prime\prime}$, and 1.67$^{\prime\prime}$, and correspond to sources at $z_s=1.54,$ $1.07$, $2.11$, respectively. In Section 2, we describe the two selection techniques, introduced by \citet{Wil++17} and \citet{Agn++17}, and their application to SDSS data. In Section 3, we present the follow-up imaging and spectroscopy observations of the candidates and provide simple model fits to the three lenses. Finally, we conclude in Section 4. 

\section{Candidate selection methods}
\label{sect:selection}
The colours of lensed quasars are a combination of the colours of quasars and those of the lensing galaxy. This places them in a particular location in colour-magnitude space that is separate from the locations of more common contaminant classes of objects such as individual quasars or individual galaxies. We use two independent selection techniques that take advantage of this fact: one which describes the distribution of all classes of objects individually, and another which uses pseudo-distance measures in colour-magnitude space to identify objects that lie `far' away from more common classes/clusters.

\subsection{Population Mixture Models}
\label{sect:popmix}
The population mixture model approach attempts to describe the populations of lensed quasars and various contaminant classes as a superposition of $K$ probability density functions (PDFs) in colour-magnitude space. We use Gaussian PDFs and the Expectation Maximization algorithm to fit the Gaussians to the data, where each Gaussian is associated with a different class of objects. This then allows us to compute the probability that a particular object belongs to each of the $K$ classes.

The objects in this paper were selected using the results of three different models utilizing SDSS and WISE photometry: a 6 feature model including $g_\text{mod} - r_\text{mod}$, $g_\text{mod} - i_\text{mod}$, $r_\text{mod} - z_\text{mod}$, $i_\text{mod} - W1$, $W1 - W2$, and $W2$; a 7 feature model adding $W2 - W3$; and a 9 feature model adding $(g_\text{psf} - g_\text{mod}) - (r_\text{psf} - r_\text{mod})$, $(r_\text{psf} - r_\text{mod}) - (i_\text{psf} - i_\text{mod})$, and $(i_\text{psf} - i_\text{mod}) - (z_\text{psf} - z_\text{mod})$ as a measure of extendedness. Objects were first selected from the SDSS DR12 data set according to the cuts in \citep{Wil++17} and were then run through each of the 6, 7, and 9 feature models, generating three membership probability vectors for each object. We retained only those objects where the average `lens' probability across the three models is greater than 0.8. Each of these was visually inspected by two investigators (among PW, AA, TT) using the SDSS DR12 Image List Tool and assigned a score of 0-3 with the following grading scheme: 0 - not a lens, 1 - probably not a lens, 2 - possibly a lens, 3 - probably a lens. Those receiving an average visual inspection score greater than 2 were then selected for follow-up based on observability. 

\subsection{Outlier selection}
\label{sect:blobsel}

In the outlier selection technique, Gaussians are used to characterize four classes, or \textit{clusters}, of `common' objects: nearby quasars ($z < 0.75$), isolated quasars at higher redshift, blue-cloud galaxies, and faint objects. Each Gaussian $k$ is characterized by a mean $\bm{\mu}_k$ and a covariance ${\bf C}_k$ in the seven-dimensional space of $g - r$, $g - i$, $r - z$, $i - W1$, $W1 - W2$, $W2 - W3$, and $W2$. The six- and nine-dimensional spaces of the population mixture model approach are not used. For a given object, a set of four pseudo-distances $d_k = 0.5\langle {\bf f} - \bm{\mu}_k, {\bf C}_k^{-1}({\bf f} - \bm{\mu}_k)\rangle$ can be calculated, describing how close the object colours are to those of the main clusters of objects. By excluding those with distances less than a certain threshold, one retains mainly peculiar objects, including lensed quasars.

\begin{figure*}
\centering
\includegraphics[width=2.1in]{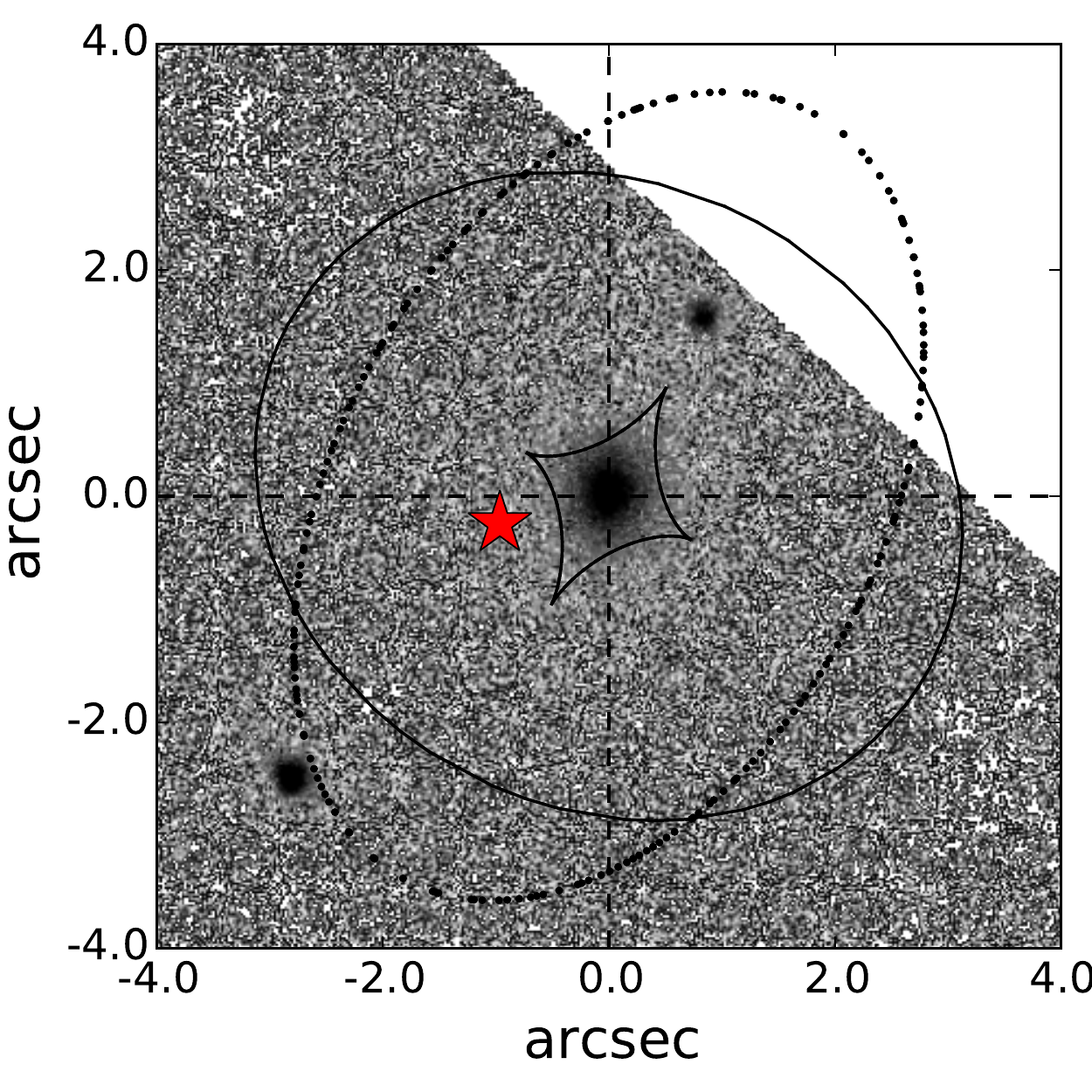}
\includegraphics[width=2.1in]{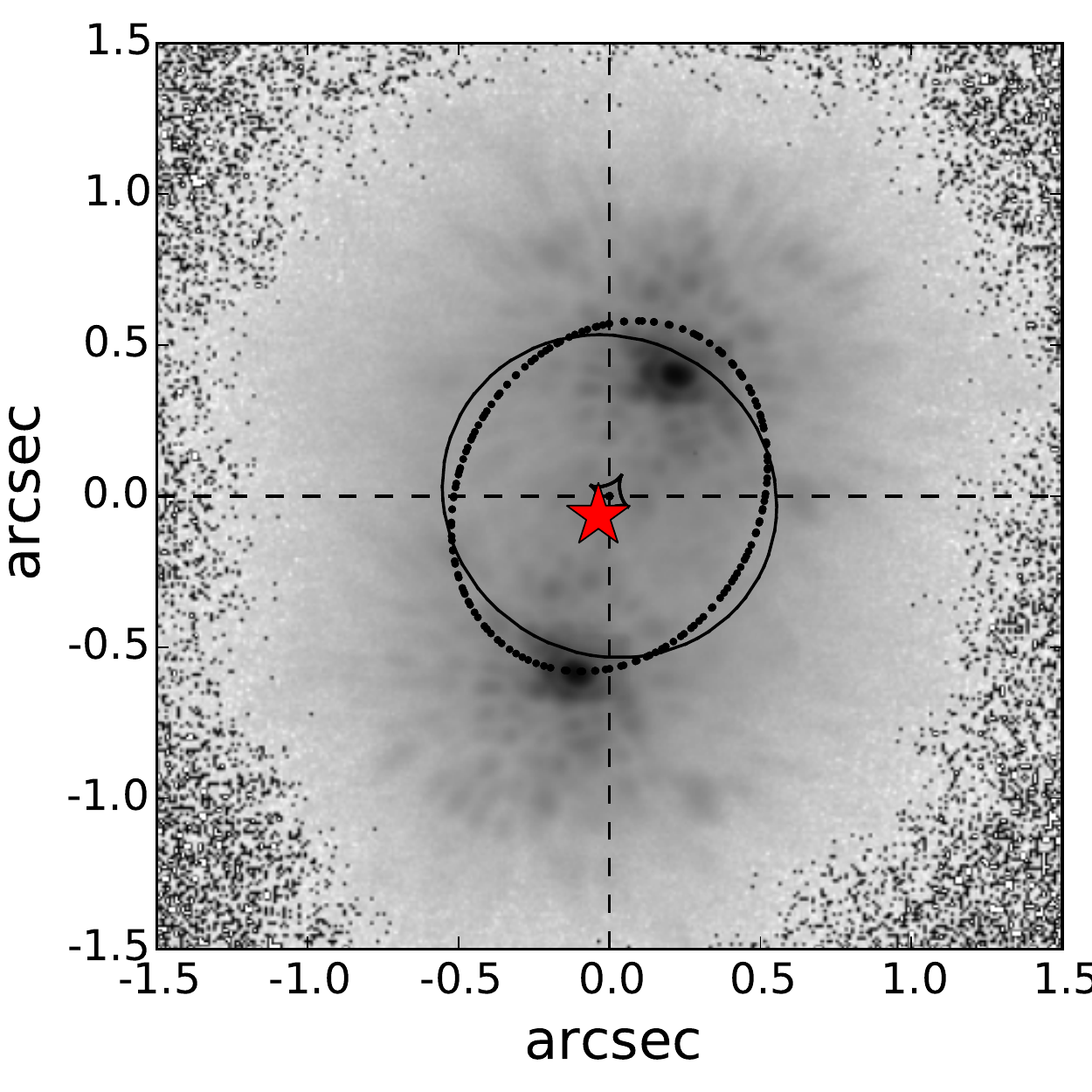}
\includegraphics[width=2.1in]{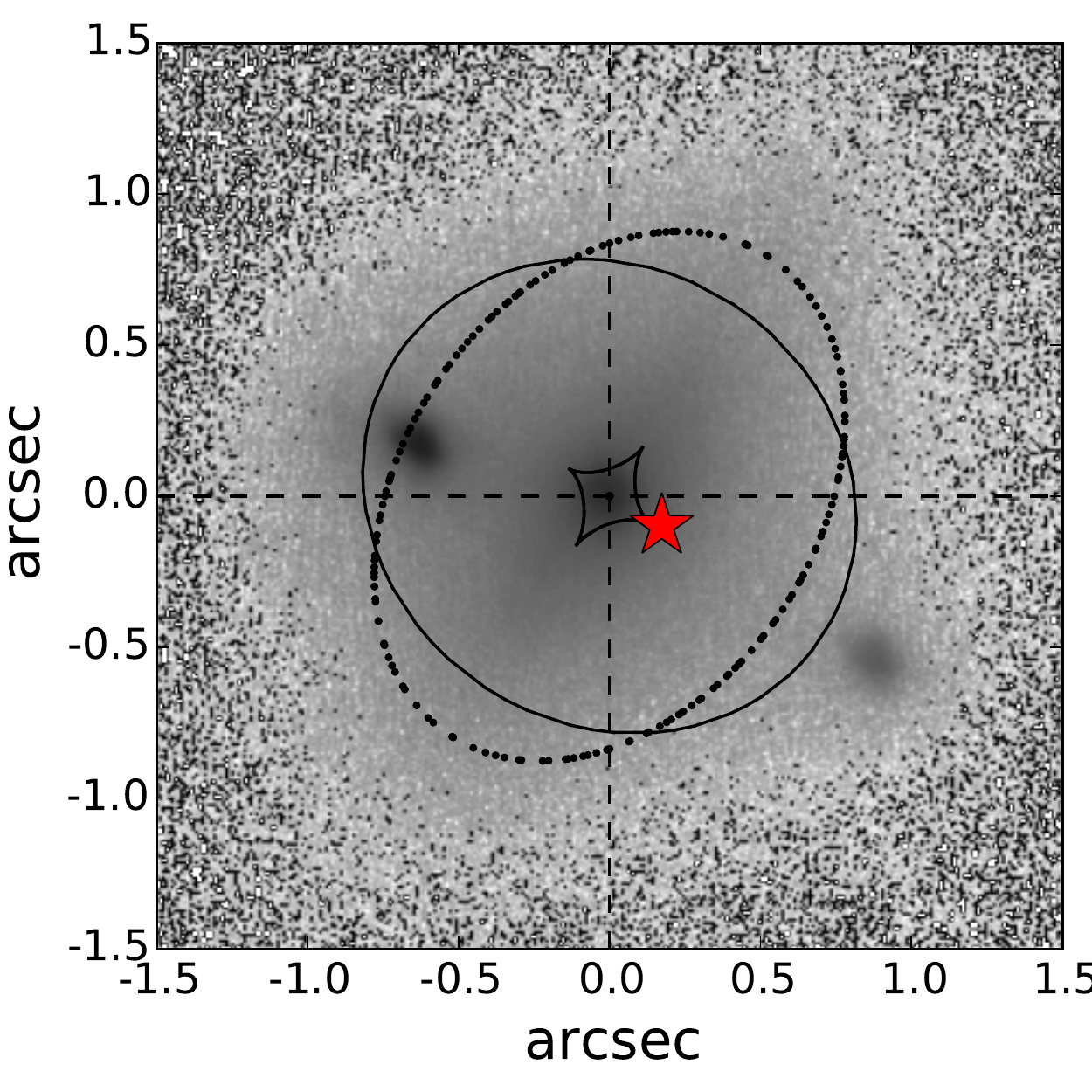}
\includegraphics[width=2.1in]{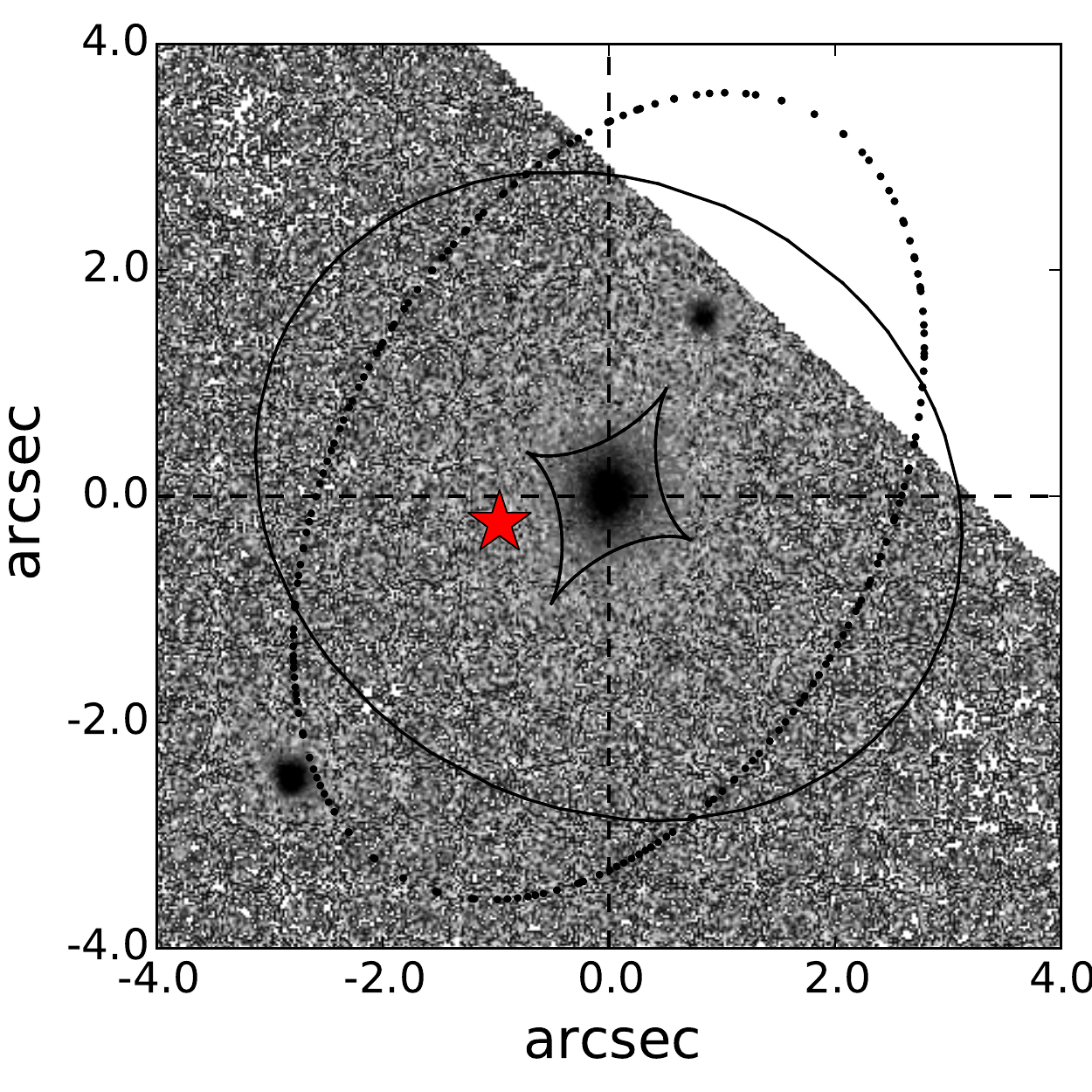}
\includegraphics[width=2.1in]{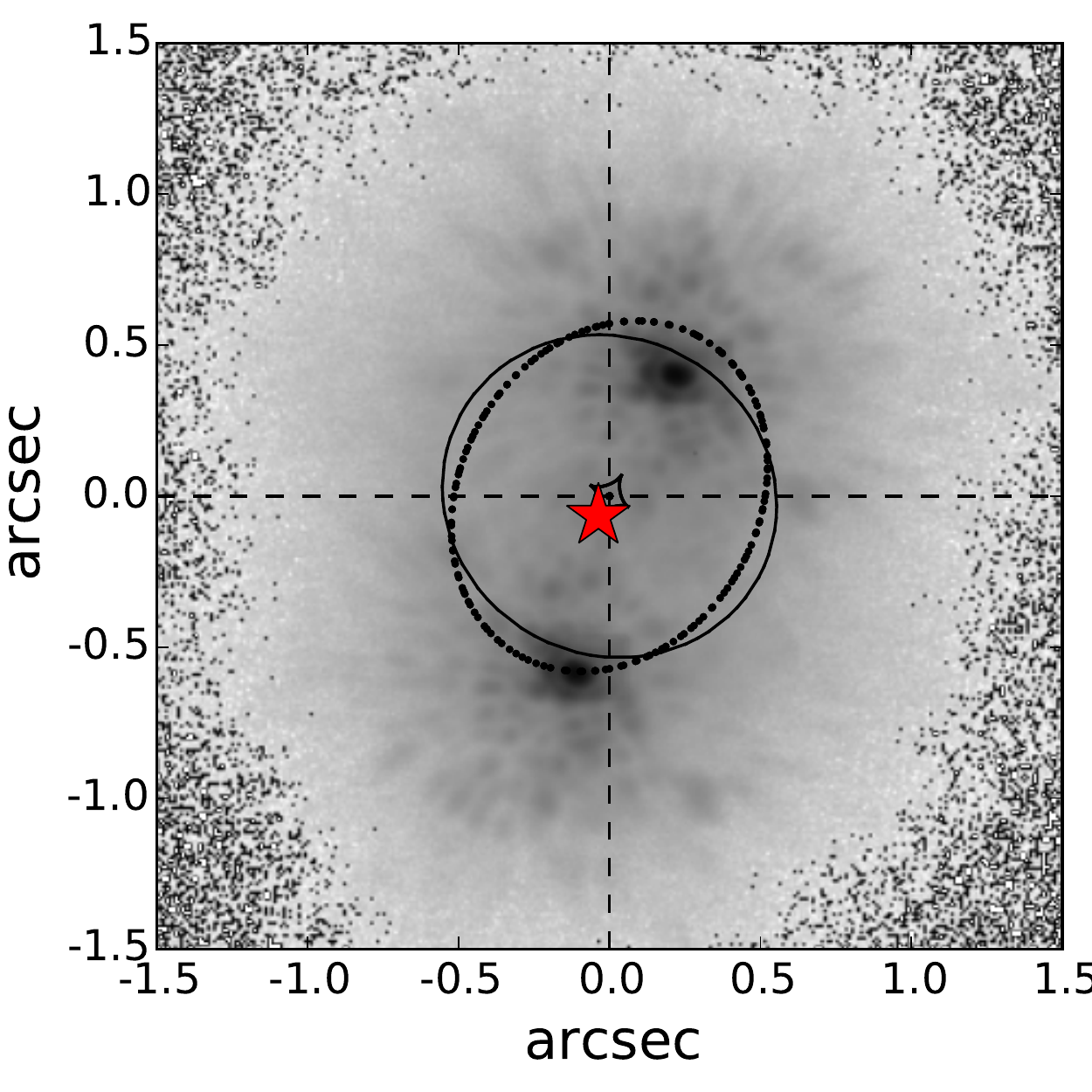}
\includegraphics[width=2.1in]{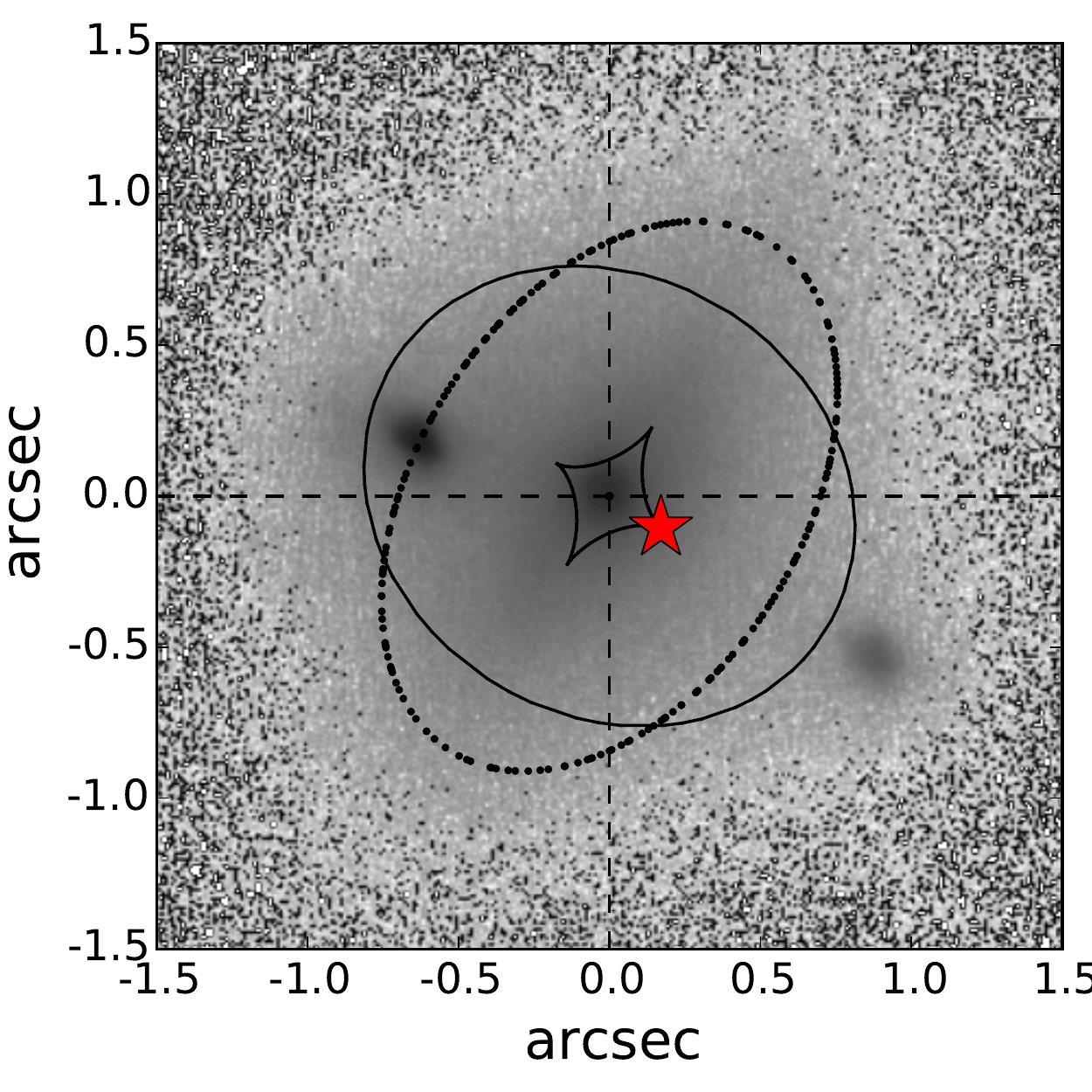}
\caption{Images of the three confirmed lenses with critical lines (dotted) and caustics (solid) overlaid. The red star-symbol indicates the location of the source. All images are aligned with North up and East left. {\it Left}: NIRC2 image of SDSS J0941+0518 made up of 27 exposures of 30 sec each, giving a total exposure time of 13.5 minutes. The image separation is $5.46^{\prime\prime}$. The lens models are SIE ({\it top}) and SIE + fluxes ({\it bottom}). {\it Middle}: NIRC2 image of SDSS J2211+1929 made up of nine 120 second exposures and two 60 second exposures, giving a total exposure time of 20 minutes. The image separation is $1.04^{\prime\prime}$. We associate the slight excess between the two point sources with the lens galaxy. The lens models are SIE ({\it top}) and SIE + fluxes ({\it bottom}). {\it Right}: OSIRIS image of SDSS J2257+2349 made up of six 120 second exposures and two 60 second exposures, giving a total exposure time of 14 minutes. Two point sources are visible at either side of an E/S0 galaxy, separated by $1.67^{\prime\prime}$. The lens models are SIE ({\it top}) and SIE + fluxes ({\it bottom}).}
\label{fig:images}
\end{figure*}

Objects were pre-selected to have extended morphology based on the $\log_{10}\mathcal{L}_{star,i}$ and \texttt{psf-model} magnitudes in the SDSS. Additional cuts in WISE magnitudes were used to exclude most $z<0.35$ quasars and narrow-line galaxies \citep[see][for details]{Agn++17}. For the remaining objects, the distances $d_k$ were calculated and only those with large enough distances were retained. 

\subsection{Performance of the two methods}
\label{sect:popmix_results}
The population mixture model yielded 59 candidates with a score of greater than 2, including 7 already known lenses and 6 objects that had already been identified, observed, and rejected. Of the remaining 46 candidates, 21 were observed, resulting in 2 confirmed lenses, 15 rejects, and 4 inconclusive. This gives a success rate between 9/59 and 38/59, ie. at least 15\%. The two main sources of contamination were close QSO + star alignments and single point sources, accounting for 9 of the 15 observed rejects and 1 of the 6 previously rejected candidates. 

The outlier selection technique yielded $\approx$ 250 candidates with $i < 20.0$, including $\approx$40 known lenses\footnote{The exact number depends on the colour-preselection cuts and pseudo-distance cuts used. The combination used for this search retained 37 previously known lenses.}. 36 of the candidates were observed, yielding 2 confirmed lenses, 22 rejects, and 12 inconclusive. This would correspond to $\geq17\%$ success rate.

\section{Follow-up of candidates}
\label{sect:followup}
Candidates were observed using the Near-Infrared Camera 2 (NIRC2), the Echelette Spectrograph and Imager \citep[ESI, ][]{She++02}, and the Optical, Spectrographic, and Infrared Remote Imaging System \citep[OSIRIS][]{Lar++06} at Keck Observatory; the ESO Faint Object Spectrograph and Camera \citep[EFOSC2][]{Buz++84} at the New Technology Telescope (NTT); and the Adaptive Optics Module (SAM) and Goodman Spectrograph at the Southern Astrophysical Research Telescope (SOAR). The instrument setups and observing conditions are summarized in Table \ref{tab:instruments}. In total, we observed 55 candidates (results summarized in the Appendix), of which 3 were confirmed to be lenses.

\begin{table*} 
\centering
\begin{tabular}{lcccrr}											
\hline											
name	&	Image separation ($^{\prime\prime}$)	&	Image	&	Flux (arbitrary)	&	$\Delta\text{r.a.}\cdot \cos$(dec.) ($^{\prime\prime}$)	&	$\Delta$dec. ($^{\prime\prime}$)	\\
\hline											
J0941+0518	&	$5.4554 \pm 0.0003$	&	A	&	$1.49~(\pm 20\%)$	&	$-2.7998 \pm 0.0003$	&	$-2.5096 \pm 0.0003$	\\
	&		&	B	&	$1.00~(\pm 20\%)$	&	$0.8349 \pm 0.0003$	&	$1.5586 \pm 0.0003$	\\
J2211+1929	&	$1.0391 \pm 0.0001$	&	A	&	$1.20~(\pm 20\%)$	&	$0.2218 \pm 0.0005$	&	$0.3924 \pm 0.0005$	\\
	&		&	B	&	$1.00~(\pm 20\%)$	&	$-0.1080 \pm 0.0005$	&	$-0.5930 \pm 0.0005$	\\
J2257+2349	&	$1.6701 \pm 0.0036$	&	A	&	$3.85~(\pm 20\%)$	&	$-0.6219 \pm 0.0010$	&	$0.1667 \pm 0.0010$	\\
	&		&	B	&	$1.00~(\pm 20\%)$	&	$0.8873 \pm 0.0037$	&	$-0.5485 \pm 0.0037$	\\
\hline											
\end{tabular}											
\caption{Image positions and fluxes used to fit the lens models. Positions are measured relative to the lens galaxy.}
\label{tab:lens_positionsandfluxes}
\end{table*}

\begin{table*} 
\centering
\begin{tabular}{lccccc}											
\hline											
name	&	Lens model	&	$b$ ($^{\prime\prime}$)	&	$e$	&	$\theta_e$ (deg)	&	$\chi^2$	\\
\hline											
J0941+0518	&	SIE	&	$2.88_{-0.66}^{+0.01}$	&	$0.37_{-0.21}^{+0.01}$	&	$-28_{-1}^{+70}$	&	$5.37\times 10^{-2}$	\\
	&	SIE + fluxes	&	$2.88_{-0.01}^{+0.01}$	&	$0.37_{-0.01}^{+0.01}$	&	$-28.2_{-0.9}^{+0.9}$	&	$30.3$	\\
J2211+1929	&	SIE	&	$0.541_{-0.04}^{+0.001}$	&	$0.186_{-0.12}^{+0.003}$	&	$-29.8_{-70.6}^{+0.3}$	&	$4.1\times 10^{-3}$	\\
	&	SIE + fluxes	&	$0.541_{-0.001}^{+0.001}$	&	$0.186_{-0.009}^{+0.001}$	&	$-29.7_{-0.7}^{+0.3}$	&	$27.3$	\\
J2257+2349	&	SIE	&	$0.79_{-0.03}^{+0.09}$	&	$0.28_{-0.18}^{+0.08}$	&	$-34^{+2}$	&	$2.0\times 10^{-4}$	\\
	&	SIE + fluxes	&	$0.761_{-0.001}^{+0.001}$	&	$0.357_{-0.003}^{+0.002}$	&	$-31.7_{-0.01}^{+0.01}$	&	$33.6$	\\
\hline											
\end{tabular}											
\caption{List of confirmed lenses and their best-fit model parameters. The 1-$\sigma$ error is found by determining the parameters for which $\chi^2 = \chi_{\text{min}}^2 + 1$. The angle $\theta_e$ is measured in degrees East of North.}
\label{tab:lens_properties}
\end{table*}

We used the \texttt{lensmodel} package \citep{Kee11} to fit simple lens models to each confirmed lens. We adopted a singular isothermal ellipsoid (SIE) model, which has a surface density given by
\begin{align}
\kappa = \frac{b}{2\sqrt{(1-\epsilon)x^2 + (1+\epsilon)y^2}},
\end{align}
where $b$ is the lens strength, and the axis ratio $q$ is related to $\epsilon$ by $q^2 = (1 - \epsilon)/(1 +\epsilon)$. We use the \texttt{lensmodel} routine \texttt{optimize} which minimizes $\chi^2$ by varying the source and galaxy positions, the source flux, the lens strength $b$, the ellipticity $e = 1 - q$, and the ellipticity position angle $\theta_e$. Our observational constraints are the image and galaxy positions found using a centroid algorithm and fluxes found using aperture photometry with the high resolution AO images. Since flux ratios can be affected by microlensing and differential extinction \citep[e.g.,][]{M+S98,Fal++99} as well as the combined effects of quasar variability and time delays, the flux measurements introduce additional uncertainties greater than those measured. For this reason, we fit a model without using image fluxes and one using the fluxes with 20\% uncertainties. The relative positions and fluxes of the images are given in Table \ref{tab:lens_positionsandfluxes}. With the fluxes omitted, we have 4 constraints from the image positions relative to the galaxy and are fitting for $b$, $e$, $\theta_e$, and source position. This means we have more free parameters than constraints and should expect a $\chi^2\sim 0$. In this case, the fitted parameters give an idea of the correct values for the SIE model, but the errors do not hold as much meaning as those for the fits including the flux ratios. The best-fit parameters for all models are given in Table \ref{tab:lens_properties}.

\subsection{SDSS J0941+0518}
\label{sect:J0941}
Images of SDSS J0941+0518 were obtained with the OSIRIS Imager at Keck Observatory on 9 November 2016. The Kbb filter was used, centered at 2172~nm and with bandwith 415~nm, with a scale of 0.02 arcsec/pixel. 27 exposures of 30 seconds each were obtained with 3$^{\prime\prime}$ dithers in each direction. Spectra were obtained with ESI at Keck Observatory on 19 November 2016, in echelette mode with a wavelength coverage from 3900 \AA ~to 10900 \AA ~and dispersion ranging from 0.16 \AA/pixel to 0.30 \AA/pixel, giving a constant dispersion of 11.5 km/sec/pixel. The 0.75 arcsec slit was used, which projects to 4.9 pixels. Three 600s exposures were taken, dithering along the slit between exposures.

\begin{figure*}
\centering
\includegraphics[height=2.15in]{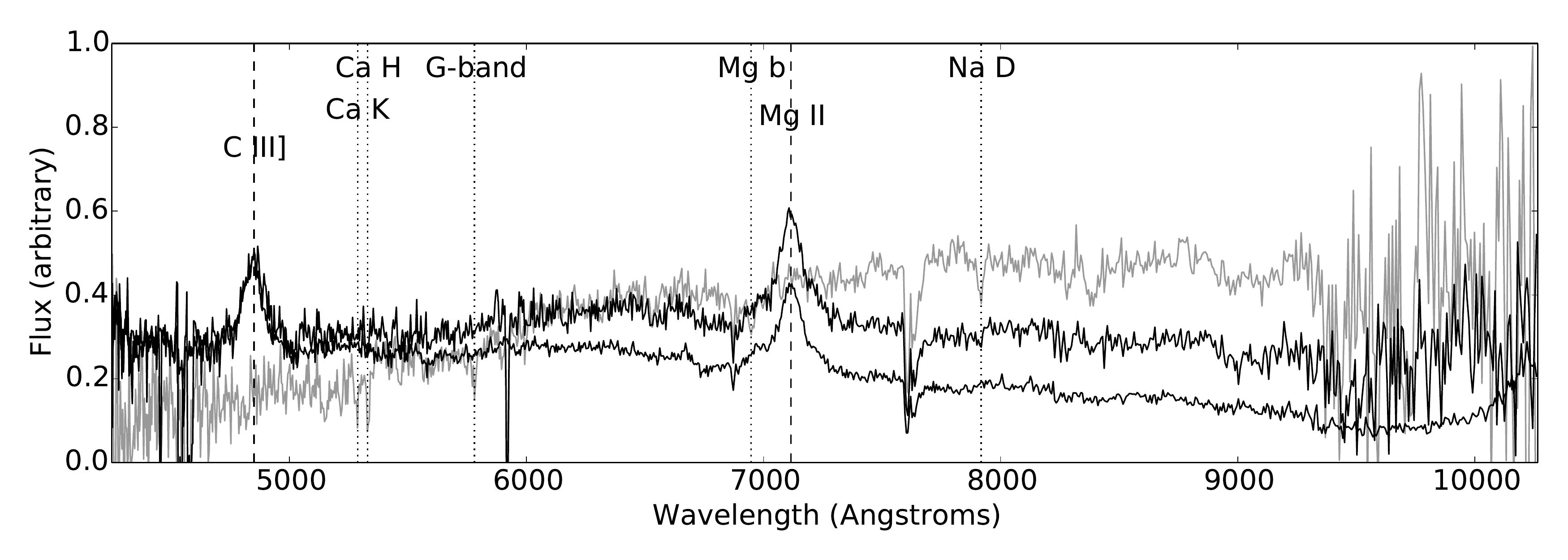}
\includegraphics[height=2.15in]{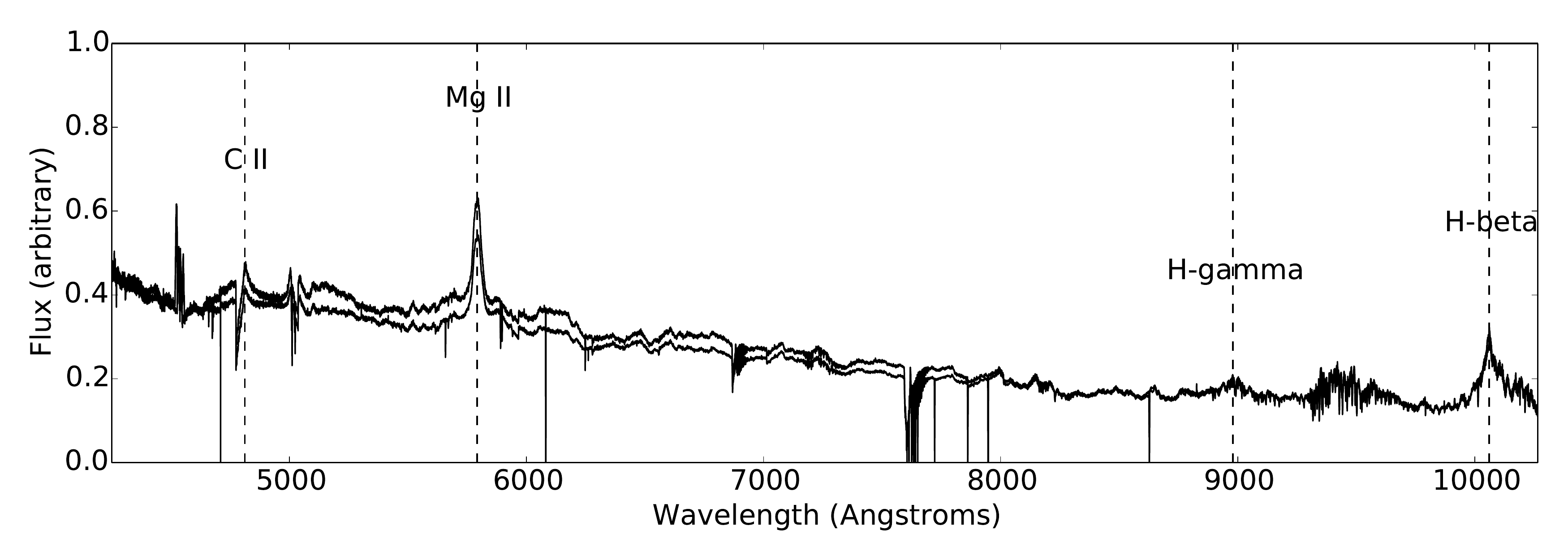}
\includegraphics[height=2.15in]{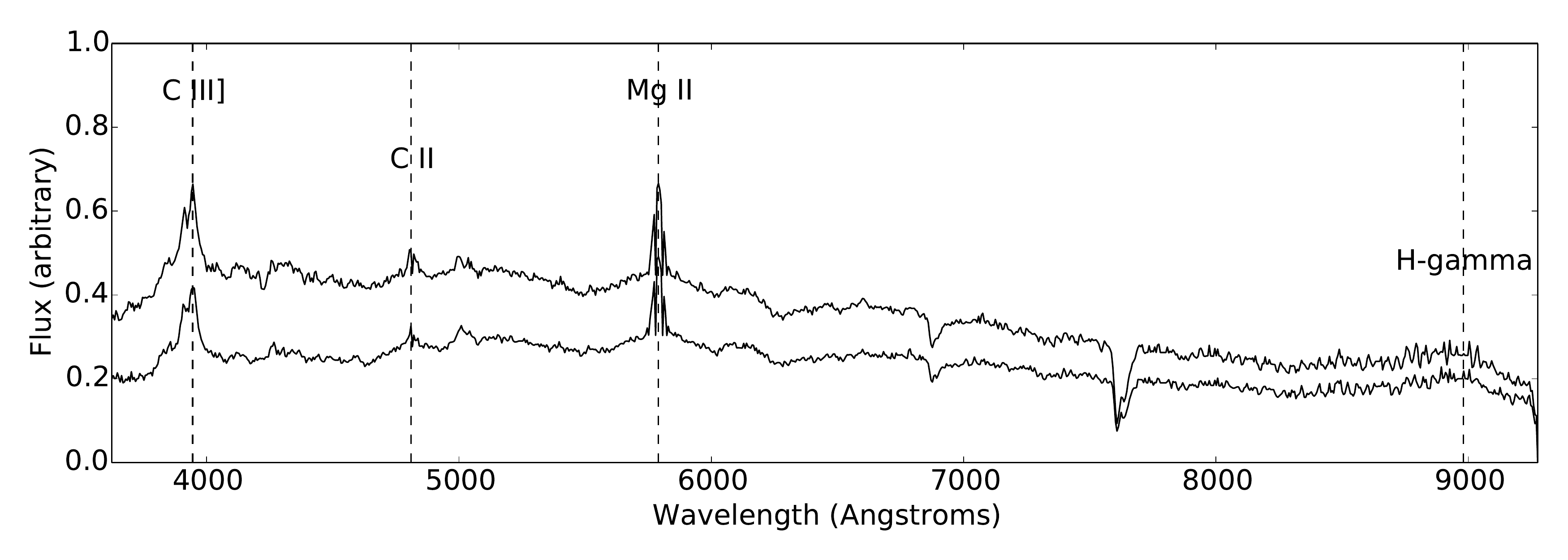}
\includegraphics[height=2.15in]{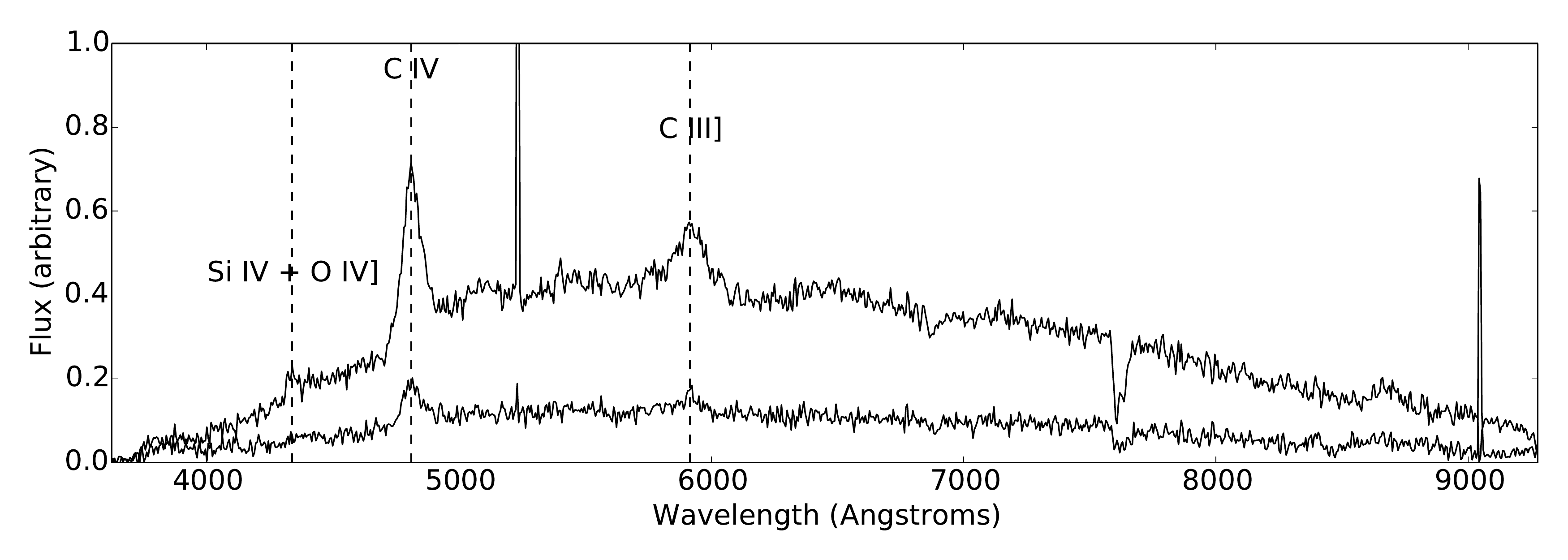}
\caption{Extracted 1D spectra of the quasar images in the three confirmed lenses. Identifiable emission lines are marked with vertical dashed lines, and absorption features are marked with dotted lines. Note that the features at 7600-7630\AA~and 6860-6890\AA~are the A-band and B-band atmospheric absorption features and should not be attributed to the quasar spectra. From top to bottom, the panels are (1) SDSS J0941+0518, ESI; (2) SDSS J2211+1929, ESI; (3) SDSS J2211+1929, EFOSC2; (4) SDSS J2257+2349, EFOSC2. In panel (1), the galaxy spectrum is plotted in gray. Note that one of the quasar spectra is contaminated by the galaxy spectrum, as can be seen by the excess and similar features at the red end.}
\label{fig:spectra}
\end{figure*}

Images show two point sources separated by 5.46$^{\prime\prime}$ on either side of the lens galaxy. The spectra (Figure \ref{fig:spectra}, top panel) are consistent with coming from the same object at redshift $z_s = 1.54$, as identified by the C III] and Mg II emission lines. The extracted galaxy spectrum with the Ca H and K, G-band, Mg b, and Na D lines give $z_l = 0.343$. These values agree with the SDSS fiber spectra which give $z_s = 1.55$ and $z_l = 0.343$. 

The SIE model fit infers an elliptical lensing object with ellipticity $e = 0.37\pm 0.01$ and position angle $\theta_e = -28.2\pm 0.9$ degrees, measured East of North. The inferred mass centroid from the model agrees with the light centroid from the images. The resulting critical lines and caustics are plotted on the images in Figure \ref{fig:images} along with the inferred source position. The $\chi^2$ value for the model without flux constraints was $5.37\times 10^{-2}$, but the ellipticity and position angle were poorly constrained. Including the fluxes increased $\chi^2$ to 30.3, but the parameter constraints improved drastically.

The lens galaxy is in close proximity ($<0.5^\prime$) to eight other sources in SDSS imaging. \citet{Con++09} identify 6 of these objects, including the lens, as belonging to a compact group of galaxies. This could explain the large image separation and indicates the need for a more complex lens model.

\begin{figure}
\centering
\includegraphics[width=0.45\textwidth]{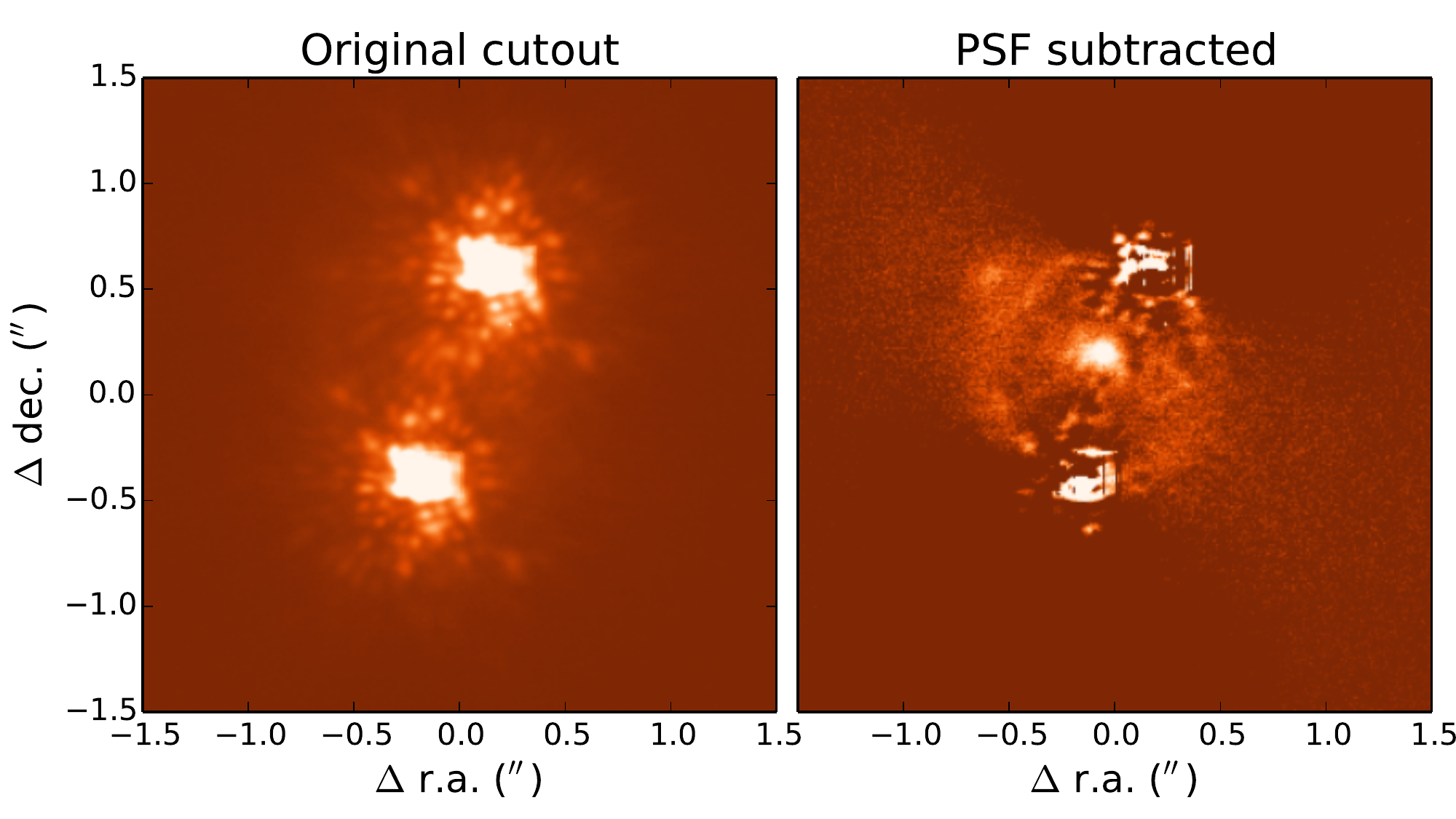}
\caption{A simple PSF subtraction reveals the lens galaxy and and Einstein ring in the NIRC2 AO-assisted image of SDSS J2211+1929. The PSF subtraction was performed by subtracting a flux scaled cutout of Image A from Image B and vice versa. Note that since the two image peaks do not lie on pixel centers, the subtraction is not perfectly aligned.}
\label{fig:2211_psfsubtracted}
\end{figure}

\subsection{SDSS J2211+1929}
\label{sect:J2211}
Images of SDSS J2211+1929 were obtained using NIRC2 with adaptive optics at Keck Observatory on 21 September 2016. The narrow camera was used with the Kp filter, centered at 2124 nm and with bandpass 351 nm. The pixel scale is 0.01 arcsec/pixel. Three sets of three 120s exposures were taken using a three point dither pattern in addition to two 60s acquisition exposures. Spectra were first obtained with ESI at Keck Observatory on 19 November 2016. The same setup was used as for SDSS J0941+0518, but with two 900s exposures. A second set of spectra were obtained with EFOSC2 at NTT on 27 September 2016. The Gr\#13 grism was used with a $1.2^{\prime\prime}$ slit, which covers 3685 \AA ~to 9315 \AA ~with dispersion 5.54 \AA/pixel.

The images show two bright point sources separated by 1.04$^{\prime\prime}$. Despite the complex structure in the PSF, a slight excess can be seen which we associate with the lens galaxy (Figure \ref{fig:2211_psfsubtracted}, left). A simple pairwise PSF subtraction, in which Image A is subtracted from Image B and vice versa, reveals the lens galaxy as well as an Einstein ring (Figure \ref{fig:2211_psfsubtracted}, right). The extracted 1D spectra are consistent with coming from the same object with emission lines C III], C II, Mg II, H-gamma, and H-beta at redshift $z_s = 1.07$.

The SIE fit with fluxes infers an ellipticity $e = 0.186_{-0.009}^{+0.001}$ and position angle $\theta_e = -29.7_{-0.7}^{+0.3}$. As with J0941+0518, the fit without flux observations does not reliably constrain the ellipticity position angle, but including fluxes with $20\%$ uncertainties significantly improves the constraints. The inferred mass centroid from the SIE fit again agrees with the light centroid from the psf-subtracted images.

\subsection{SDSS J2257+2349}
\label{sect:J2257}
Images of SDSS J2257+2349 were obtained with NIRC2 with adaptive optics at Keck Observatory on 21 September 2016. The same setup was used as for SDSS J2211+1929, but with two sets of three 120s exposures and the two additional 60s acquisition exposures. Spectra were obtained with EFOSC2 at NTT on 27 September 2016 with the same setup as for SDSS J2211+1929.

The images show two point sources separated by 1.67$^{\prime\prime}$ on either side of an E/S0 galaxy. Spectra show that both point sources correspond to the same quasar at redshift $z_s = 2.11$, as indicated by the Si IV + O IV], C IV, and C III] emission lines. 

The SIE fit with fluxes infers an ellipticity $e = 0.357_{-0.003}^{+0.002}$ and ellipticity position angle $-31.7\pm 0.01$ degrees. These values agree with the orientation of the lens galaxy as seen in the right panels of Figure \ref{fig:images} and the mass centroid from the fit matches the light centroid from the images. When fluxes are not included as constraints, the position angle is only constrained with an upper bound, but when the fluxes are used, all parameters are well constrained. This again illustrates the necessity of deeper data in order to constrain the deflector shape and explore more complex lens models. 

\section{Discussion and Conclusions}
\label{sect:concl}

We have followed-up 55 lensed quasar candidates selected by two photometry-based selection techniques, independently applied to the SDSS-DR12 data set, confirming three new lenses: J0941+0518, J2211+1929, and J2257+2349. Of these, J0941 was selected by both methods, J2211 was found only by the population mixture search, and J2257 only by the outlier selection. This reflect a general behaviour on the larger sample of selected candidates, where the two searches complement one another and have some degree of overlap.

Adaptive-Optics assisted images of the three lenses, taken with OSIRIS and NIRC2, reveal two quasar images on either side of a lensing source, separated by $5.46^{\prime\prime}$, $1.04^{\prime\prime}$, and $1.67^{\prime\prime}$, respectively. Spectra taken with ESI and EFOSC2 confirm that each system is a lens with quasar redshifts $z_s = 1.54$, $1.07$, and $2.11$, respectively, and give a lens redshift $z_l = 0.343$ for J0941+0518. Fits to simple SIE models with and without the use of fluxes as a constraint give the lens parameters summarized in Table \ref{tab:lens_properties}. With deeper data, one could apply more complex models, e.g. including external shear terms or deviations from the SIE density law, in order to better constrain the shape of the lens.

Most of the `inconclusive' objects appear as two point sources in images, but do not show any sign of a lens galaxy. These objects will require spectra to confirm if they are images of the same quasar, or, similar to the case of J2211+1929, deeper imaging with careful PSF subtraction. In future follow-up campaigns, a quick PSF subtraction such as the one done with J2211+1929 can be used as a tool for quickly examining targets while at the telescope. 

A non-negligible subset of candidates were revealed as bright and single point-sources in follow-up imaging. Their uncertain SDSS morphology was given by CCD `blooming', which is common for bright sources and in fact can be seen also on some known quasar lenses. Subsequent Pan-STARRS images\footnote{http://ps1images.stsci.edu/cgi-bin/ps1cutouts}, not available at the time of this campaign, were much clearer at distinguishing between spurious candidates (due to blooming) and systems with truly multiple images.

The discovery of these three lenses in the SDSS demonstrates the importance of photometry-based selection techniques to complement previous searches for lensed quasars. Neither J2211+1929 nor J2257+2349 have spectra in the SDSS and thus were not explored by previous searches like the SQLS.
The case of  J0941+0518 is more surprising: despite having, coincidentally, a fibre spectrum of the bright quasar image and one of the lens galaxy and counter-image, it was missed by previous spectroscopic searches.
With new and upcoming surveys that do not have readily available spectroscopic data, both types of searches will be important in order to generate a more complete sample of lenses.

\section*{Acknowledgments}
TT acknowledges support by the Packard Foundation through a Packard Research Fellowship and by the National Science Foundation through grant AST-1450141. T. A. and Y.A. acknowledge support by proyecto FONDECYT 11130630 and by the Ministry for the Economy, Development, and Tourism’s Programa Inicativa Cient\'{i}fica Milenio through grant IC 12009, awarded to The Millennium Institute of Astrophysics (MAS). CDF acknowledges support from the NSF via AST-1312329. VM acknowledges support from Centro de Astrof\'{\i}sica de Valpara\'{\i}so. K.R. is supported by PhD fellowship FIB-UV 2015/2016 and Becas de Doctorado Nacional CONICYT 2017.  

Based on observations obtained at the Southern Astrophysical Research (SOAR) telescope, which is a joint project of the Minist\'{e}rio da Ci\^{e}ncia, Tecnologia, e Inova\c{c}\~{a}o (MCTI) da Rep\'{u}blica Federativa do Brasil, the U.S. National Optical Astronomy Observatory (NOAO), the University of North Carolina at Chapel Hill (UNC), and Michigan State University (MSU).

Funding for SDSS-III has been provided by the Alfred P. Sloan Foundation, the Participating Institutions, the National Science Foundation, and the U.S. Department of Energy Office of Science. The SDSS-III web site is http://www.sdss3.org/.

SDSS-III is managed by the Astrophysical Research Consortium for the Participating Institutions of the SDSS-III Collaboration including the University of Arizona, the Brazilian Participation Group, Brookhaven National Laboratory, Carnegie Mellon University, University of Florida, the French Participation Group, the German Participation Group, Harvard University, the Instituto de Astrofisica de Canarias, the Michigan State/Notre Dame/JINA Participation Group, Johns Hopkins University, Lawrence Berkeley National Laboratory, Max Planck Institute for Astrophysics, Max Planck Institute for Extraterrestrial Physics, New Mexico State University, New York University, Ohio State University, Pennsylvania State University, University of Portsmouth, Princeton University, the Spanish Participation Group, University of Tokyo, University of Utah, Vanderbilt University, University of Virginia, University of Washington, and Yale University. 

Some of the data presented herein were obtained at the W.M. Keck Observatory, which is operated as a scientific partnership among the California Institute of Technology, the University of California and the National Aeronautics and Space Administration. The Observatory was made possible by the generous financial support of the W.M. Keck Foundation. The authors wish to recognize and acknowledge the very significant cultural role and reverence that the summit of Mauna Kea has always had within the indigenous Hawaiian community. We are most fortunate to have the opportunity to conduct observations from this mountain, and we respectfully say mahalo.

\nocite{Astropy2013}
\bibliographystyle{mnras}
\bibliography{references} 

\appendix
\section{Observed targets}
\label{sect:observed}
In Table \ref{tab:observed}, we list all 55 observed candidates and their follow-up outcomes. Of the 16 inconclusive objects, most appear as two point sources in imaging. These need either deeper imaging to bring out the lens galaxy, as in the case of J2211+1929, or spectra to confirm that both images are the same quasar. The ruled out candidates are split into those that are single point sources and those that are multiple sources, but not lenses. The single-source objects appear as single point sources in the SDSS, but were selected as potential small separation lenses because they contain galaxy colours and some have SDSS QSO spectra showing galaxy absorption features. These objects can probably be avoided in future searches without risk of losing a substantial number of lenses.

\newcommand*\rot{\rotatebox{90}}
\begin{sidewaystable*} 
\renewcommand{\arraystretch}{0.85}
\centering
\small
\begin{tabular}{llrrccclcc}																			
\hline																			
	&	name	&	r.a.(J2000)	&	dec.(J2000)	&	mag\_i	&	selection method	&	Instrument	&	Comments	&	$z_s$	&	$z_l$	\\
\hline																			
	&	J0941+0518	&	145.3435481	&	5.3069880	&	21.39, 17.44, 18,51	&	PopMix, OutlierSel	&	ESI, OSIRIS	&		&	1.54	&	0.343	\\
	&	J2211+1929	&	332.8763774	&	19.4869926	&	15.41	&	PopMix	&	EFOSC2, ESI, NIRC2	&		&	1.07	&		\\
\rot{\rlap{Lens}}	&	J2257+2349	&	344.3558623	&	23.8251034	&	17.67	&	OutlierSel	&	EFOSC2, NIRC2	&		&	2.11	&		\\
																			\\
	&	J0048+2505	&	12.1457148	&	25.0896541	&	18.77	&	OutlierSel	&	NIRC2	&	two point sources	&		&		\\
	&	J0118+4718	&	19.7397002	&	47.3147867	&	18.21	&	OutlierSel	&	NIRC2	&	two point sources	&		&		\\
	&	J0130+4110	&	22.5096548	&	41.1693817	&	18.22	&	OutlierSel	&	NIRC2	&	two point sources	&		&		\\
	&	J0213+1306	&	33.3262661	&	13.1121799	&	18.47	&	OutlierSel	&	NIRC2	&	bad AO correction	&		&		\\
	&	J0252+3420	&	43.0729962	&	34.3382372	&	16.16, 15.96	&	OutlierSel	&	NIRC2	&	two point sources	&		&		\\
	&	J0252$-$0855	&	43.0879945	&	-8.9210091	&	17.79	&	OutlierSel	&	NIRC2	&	two point sources	&		&		\\
	&	J0852$-$0148	&	133.2245838	&	-1.8139580	&	19.42, 18.59	&	PopMix	&	ESI, SAM	&	two point sources	&		&		\\
	&	J0930+4614	&	142.5881738	&	46.2396842	&	18.25	&	OutlierSel	&	ESI	&	two traces visible, but too faint to confirm	&	2.38	&		\\
	&	J0932+0722	&	143.0298040	&	7.3809231	&	18.94	&	OutlierSel	&	EFOSC2, SAM	&	two point sources	&	1.99	&		\\
	&	J1010+5705	&	152.7130546	&	57.0919030	&	16.93	&	PopMix	&	ESI	&	one obvious, possibly two blended traces	&	1.97	&		\\
\rot{\rlap{Inconclusive}}	&	J1013+1041	&	153.4169237	&	10.6877941	&	18.44, 17.39	&	PopMix	&	SAM	&	two point sources	&		&		\\
	&	J1700+0058	&	255.1000485	&	0.9708746	&	16.13	&	OutlierSel	&	NIRC2	&	two point sources	&		&		\\
	&	J2103+1100	&	315.8419650	&	11.0053179	&	18.82	&	OutlierSel	&	NIRC2	&	two point sources	&		&		\\
	&	J2209+0045	&	332.2788165	&	0.7621817	&	19.50, 19.78	&	OutlierSel	&	NIRC2, EFOSC2, SAM	&	two point sources	&		&		\\
	&	J2246+3118	&	341.6917100	&	31.3047196	&	19.46, 20.51	&	OutlierSel	&	NIRC2	&	two point sources	&		&		\\
	&	J2352+0105	&	358.1586989	&	1.0978733	&	17.16	&	PopMix	&	ESI	&	one obvious trace	&	2.15	&	0.83?	\\
																			\\
	&	J2353$-$0539	&	358.4625516	&	-5.6655170	&	18.11, 18.40, 16.52	&	PopMix	&	NIRC2, EFOSC2	&	QSO + star	&		&		\\
	&	J0037+0111	&	9.3326860	&	1.1874153	&	18.21	&	OutlierSel	&	NIRC2	&	point source + galaxy	&		&		\\
	&	J0120+2654	&	20.0190333	&	26.9153290	&	17.15	&	OutlierSel	&	NIRC2	&	point source + galaxy	&		&		\\
	&	J0141+0007	&	25.4609249	&	0.1317267	&	19.06	&	PopMix	&	EFOSC2, ESI, SAM, NIRC2	&	QSO + galaxy, $z_\text{gal} = 0.279$	&	1.35	&		\\
	&	J0255$-$0051	&	43.9399327	&	-0.8650315	&	16.79, 19.12	&	PopMix	&	NIRC2	&	large separation, no sign of lens	&		&		\\
	&	J0739+1350	&	114.9568004	&	13.8366667	&	16.85, 18.83	&	PopMix	&	ESI	&	QSO + star	&		&		\\
	&	J0806$-$0135	&	121.6793365	&	-1.5952567	&	18.22, 17.00	&	PopMix	&	ESI	&	QSO + star	&		&		\\
	&	J0808$-$0051	&	122.0369752	&	-0.8643357	&	18.52, 16.60	&	PopMix	&	ESI	&	QSO + star	&		&		\\
	&	J0808+0118	&	122.1946461	&	1.3110056	&	19.62, 16.03	&	PopMix	&	Goodman, SAM, SAM	&	3 stars	&		&		\\
	&	J0836+4841	&	129.2064862	&	48.6983237	&	18.11, 19.03	&	OutlierSel	&	ESI	&	QSO + AGN pair	&		&		\\
\rot{\rlap{Not a lens}}	&	J0841+0312	&	130.2782996	&	3.2019064	&	16.04, 18.05	&	PopMix	&	ESI	&	QSO + star	&		&		\\
	&	J0940$-$0249	&	145.1593119	&	-2.8267049	&	17.28	&	PopMix	&	ESI	&	$z \approx 0.092$	&		&		\\
	&	J1704+1817	&	256.1354746	&	18.2910172	&	18.36, 20.19, 18.88	&	OutlierSel	&	NIRC2	&	three point sources	&		&		\\
	&	J1810+6344	&	272.5184057	&	63.7407189	&	18.99, 18.52	&	PopMix	&	NIRC2	&	point source + galaxy	&		&		\\
	&	J2036$-$1801	&	309.2195540	&	-18.0292749	&	18.04, 17.69	&	OutlierSel	&	NIRC2, EFOSC2	&	QSO + star	&	2.32	&		\\
	&	J2044+0314	&	311.2035735	&	3.2486361	&	17.22	&	PopMix	&	NIRC2	&	three souces in wrong configuration	&		&		\\
	&	J2055$-$0515	&	313.8753026	&	-5.2504530	&	18.46, 19.01	&	OutlierSel	&	NIRC2, EFOSC2	&	QSO + star	&		&		\\
	&	J2350$-$0749	&	357.5107284	&	-7.8258943	&	18.18, 18.57	&	OutlierSel	&	NIRC2, EFOSC2	&	star forming galaxy	&		&		\\
																			\\
	&	J0001+1411	&	0.3166458	&	14.1897373	&	18.49	&	OutlierSel	&	NIRC2	&	SDSS - QSO	&		&		\\
	&	J0005+2031	&	1.4974794	&	20.5235499	&	17.10	&	OutlierSel	&	NIRC2	&		&		&		\\
	&	J0024+0032	&	6.1838018	&	0.5393054	&	17.02	&	PopMix, OutlierSel	&	NIRC2	&	SDSS - QSO	&		&		\\
	&	J0116+4241	&	19.0677562	&	42.6953976	&	16.86	&	OutlierSel	&	NIRC2	&		&		&		\\
	&	J0128+0055	&	22.0456250	&	0.9317422	&	17.88	&	OutlierSel	&	NIRC2	&	SDSS - QSO	&		&		\\
	&	J0209$-$0028	&	32.4458329	&	-0.4781211	&	18.86	&	PopMix	&	EFOSC2, SAM	&	QSO	&	1.31	&		\\
	&	J0242+0057	&	40.6679989	&	0.9575364	&	16.59	&	OutlierSel	&	NIRC2	&	SDSS - QSO	&		&		\\
	&	J0340+0057	&	55.1983256	&	0.9599661	&	17.81	&	OutlierSel	&	NIRC2	&		&		&		\\
	&	J0502+1310	&	75.6155639	&	13.1822185	&	18.33	&	OutlierSel	&	NIRC2	&		&		&		\\
	&	J1738+3222	&	264.7017777	&	32.3767753	&	17.41	&	OutlierSel	&	NIRC2	&		&		&		\\
	&	J2045$-$0101	&	311.4023578	&	-1.0299982	&	16.42	&	PopMix	&	NIRC2	&	SDSS - QSO	&		&		\\
	&	J2111$-$0012	&	317.7877331	&	-0.2164685	&	18.38	&	OutlierSel	&	NIRC2	&	SDSS - QSO	&		&		\\
	&	J2121$-$0005	&	320.3755890	&	-0.0908685	&	17.00	&	OutlierSel	&	NIRC2	&	SDSS - QSO	&		&		\\
\rot{\rlap{Not a lens, single source}}	&	J2123$-$0050	&	320.8727805	&	-0.8480401	&	16.34	&	PopMix	&	NIRC2	&	SDSS - QSO	&		&		\\
	&	J2146+0009	&	326.5554674	&	0.1585651	&	19.74	&	OutlierSel	&	NIRC2	&	SDSS - QSO	&		&		\\
	&	J2158+1526	&	329.6736308	&	15.4374749	&	17.47	&	OutlierSel	&	NIRC2	&		&		&		\\
	&	J2238+2718	&	339.5371608	&	27.3136765	&	18.08	&	OutlierSel	&	NIRC2	&		&		&		\\
	&	J2358$-$0136	&	359.5858434	&	-1.6029082	&	16.71	&	OutlierSel	&	NIRC2	&		&		&		\\
\hline																			
\end{tabular}																		
\label{tab:observed}													
\end{sidewaystable*}

\label{lastpage}
\end{document}